\documentclass[12pt]{article}
\usepackage{a4wide}
\usepackage{amssymb}
\usepackage{graphicx}
\begin{document}
{\renewcommand{\thefootnote}{\fnsymbol{footnote}}
\begin{center}
{\LARGE  Quasiclassical solutions for static quantum black holes}\\
\vspace{1.5em}
K\"allan Berglund,$^1$\footnote{e-mail
  address: {\tt kallan$\underline{\;\;}$berglund@alumni.brown.edu}}
Martin Bojowald,$^1$\footnote{e-mail address: {\tt bojowald@psu.edu}}
 Manuel D\'{\i}az$^2$\footnote{e-mail
   address: {\tt manueldiaz@umass.edu}} and
 Gianni Sims$^3$\footnote{e-mail
  address: {\tt  gsims2021@fau.edu}} \\
\vspace{1em}
$^1$ Institute for Gravitation and the Cosmos,
The Pennsylvania State
University,\\
104 Davey Lab, University Park, PA 16802, USA\\
\vspace{0.5em}
$^2$  Amherst Center for Fundamental Interactions,\\ Department of Physics,
University of Massachusetts Amherst,\\
426 Lederle Graduate Research Tower,
Amherst, MA 01003 USA \\
\vspace{0.5em}
 $^3$ Department of Physics, Florida Atlantic University,\\
777 Glades Road, Boca Raton, FL 33431\\
\end{center}
}

\vspace{1.5em}

\setcounter{footnote}{0}
\begin{abstract}
  A new form of quasiclassical space-time dynamics for constrained systems
  reveals how quantum effects can be derived systematically from canonical
  quantization of gravitational systems. These quasiclassical methods lead to
  additional fields, representing quantum fluctuations and higher moments,
  that are coupled to the classical metric components. The new fields describe non-adiabatic
  quantum dynamics and can be interpreted as implicit formulations of
  non-local quantum corrections in a field theory. This field-theory aspect is
  studied here for the first time, applied to a gravitational system for which
  a tractable model is constructed. Static solutions for the relevant fields
  can be obtained in almost closed form. They reveal new properties of potential
  near-horizon and asymptotic effects in canonical quantum gravity and demonstrate the overall consistency of the formalism.
\end{abstract}

\section{Introduction}

For some time now, black holes have presented a popular testing ground for possible implications of quantum gravity. Examples include quantum
corrections to Newton's law, modified horizon dynamics, implications for
Hawking radiation, tools to address the information loss problem, potential resolutions of the central singularity, or speculations about the post-singular life of a black hole. A large variety of methods have been applied, ranging from
effective field theory \cite{EffectiveGR,EffectiveNewton,BurgessLivRev} to
proposed non-perturbative ingredients of approaches such as string theory or
loop quantum gravity. 

Here, we present new results using a formulation situated on the middle ground
between standard effective field theory on one hand and non-perturbative
effects on the other: We extend effective field theory by applying
non-adiabatic quantum dynamics, foregoing the derivative expansion of quantum
corrections that is implicitly assumed when they are expressed in
higher-curvature form. Our formulation will therefore be sensitive to new (and
possibly non-local) corrections, while maintaining crucial consistency conditions for constraint equations and an application to space-time physics. The importance of such consistency, related to the question of whether general covariance can be maintained by quantum corrections, has
recently been highlighted by the finding that most black-hole models or other
space-time descriptions proposed in the field of loop quantum gravity violate
covariance \cite{NonCovDressed,Disfig,BlackHoleModels}. (A more careful approach that aims to maintain covariance as much as possible has been studied in
\cite{JR,SphSymmComplex,GowdyComplex,BHSigChange,EffLine,DefSchwarzschild,DefSchwarzschild2,DefGenBH,SphSymmEff,SphSymmEff2}, using a variety of models.)
One of these no-go theorems that ruled out covariance for certain modifications encountered in models of loop quantum gravity, derived in \cite{Disfig}, relies
on the local nature of current models. 
The no-go theorem could therefore be evaded by
constructing suitable non-local quantum corrections, possibly leading to consistent implementations of modifications in covariant models. The present paper can be
considered a first step in this direction, studying non-local quantum
corrections in spherically symmetric canonical quantum gravity. We will formalize our consistency conditions in more detail when we introduce relevant ingredients of space-time physics in Section~\ref{s:SpaceTime}.

Our formulation is based on canonical methods of non-adiabatic quantum
dynamics, used for some time in various fields such as quantum chaos or
quantum chemistry
\cite{VariationalEffAc,EnvQuantumChaos,GaussianDyn,QHDTunneling} mainly for
systems with finitely many classical degrees of freedom. Related methods
\cite{CQC,CQCFields,CQCFieldsHom} have been applied recently to spherically
symmetric models of collapsing shells \cite{CQCHawking}.  Our task will be to
extend these methods to quantum field theories, and to incorporate access to
space-time structures in order to implement consistency conditions required for general covariance. In
particular, we will consider a generalization of quasiclassical methods to
constrained systems, applied to the Hamiltonian and diffeomorphism constraints
of canonical general relativity. By requiring that  quantum-corrected
constraints obey suitable Poisson brackets, known from hypersurface
deformations \cite{GenHamDyn,Katz,ADM,Regained}, we will show that such constraints can be imposed consistently and solved for modified metric components and their quantum fluctuations.

In order to reduce the complexity of these tasks, we will work with
spherically symmetric models and analyze, for now, only static solutions. In a
field theory, even static solutions are sensitive to non-adiabatic methods
because they may vary significantly in a spatial direction. An implementation
of non-adiabatic quantum dynamics with methods from other fields is therefore
of interest. In this way, we will be able to explore new quantum effects in a
tractable manner.

We will review canonical effective methods, which provide the mathematical
basis for non-adiabatic, quasiclassical dynamics in Section~\ref{s:Can}. We will first summarize the well-developed version of these methods applied to the quantum mechanics of a single degree of freedom, as well as an extension to constrained
systems. Section~\ref{s:SpaceTime} is the central part of our paper, in which
we generalize quasiclassical methods to the field theory given by spherically
symmetric gravity. We will describe the form of effective constraints
encountered in this system, and derive the equations to be solved for static
solutions with leading quasiclassical corrections. Although multiple
integrations will be required, interesting information about these solutions
can be obtained in closed form, in particular regarding the near-horizon and
the asymptotic behaviors. We will discuss the self-consistency of our solutions from the perspective of an intuitively expected behavior of quantum fluctuations, demonstrating that they are smaller in the asymptotic regime.

\section{Canonical effective theories}
\label{s:Can}

Canonical, non-adiabatic methods of quantum dynamics provide a quasiclassical
formulation in which the classical phase space, say $(q,p)$, is extended by a
certain number of quantum degrees of freedom, depending on the order in an
expansion by $\hbar$. To leading order, the classical variables $q$ and $p$
are combined with a second canonical pair, $(s,p_s)$ where $s=\Delta q$, such
that a classical potential $V(q)$ is turned into a specific effective
potential $V_{\rm eff}(q,s)$. The derivation of this effective potential (and
the physical meaning of the momentum $p_s$) requires making use of methods of Poisson geometry. 

\subsection{Effective Hamiltonians}
\label{s:EffHam}

First, if the classical system is described by a Hamiltonian
\begin{equation}
 H=\frac{p^2}{2m}+V(q)\,,
\end{equation}
one can define an effective Hamiltonian as the expectation value $H_{\rm
  eff}=\langle\hat{H}\rangle$ of the corresponding Hamilton operator, taken in
an arbitrary state. The effective Hamiltonian is therefore a function on the
state space of the system. A systematic semiclassical description
parameterizes suitable states by their expectation values of basic operators,
$q=\langle\hat{q}\rangle$ and $p=\langle\hat{p}\rangle$, as well as a series
of moments such as
$\Delta(q^n)=\langle(\hat{q}-\langle\hat{q}\rangle)^n\rangle$. Taking into
account ordering choices, we follow \cite{EffAc,Karpacz} and define a specific
set of moments of a state by
\begin{equation} \label{Delta}
 \Delta(q^np^m)=\langle(\hat{q}-\langle\hat{q}\rangle)^n
(\hat{p}-\langle\hat{p}\rangle)^m\rangle_{\rm symm}
\end{equation}
in completely symmetric (or Weyl) ordering. The moment
order, $n+m$, corresponds to the order in a semiclassical expansion,
given by $\hbar^{(n+m)/2}$. 

Moments, together with the basic expectation values, form a phase space
equipped with a Poisson bracket that is obtained by extending the definition
\begin{equation} 
 \{\langle\hat{A}\rangle,\langle\hat{B}\rangle\}=
 \frac{\langle[\hat{A},\hat{B}]\rangle}{i\hbar}
\end{equation}
using linearity and the Leibniz rule. With this bracket, the effective
Hamiltonian $H_{\rm eff}=\langle\hat{H}\rangle$ indeed generates the correct
Hamiltonian dynamics: The equation
\begin{equation} \label{Poisson}
 \{\langle\hat{A}\rangle,H_{\rm eff}\}=
 \frac{\langle[\hat{A},\hat{H}]\rangle}{i\hbar} = \frac{{\rm
     d}\langle\hat{A}\rangle}{{\rm d}t}
\end{equation}
is equivalent to quantum evolution of generic expectation values implied by
the Schr\"odinger equation. At fixed order in $\hbar$, the resulting Poisson
manifold is, in general, not symplectic. That is, it is described by a family of
symplectic leaves, on which certain Casimir functions take constant
values. A Casimir function has vanishing Poisson brackets with any other
function on the same Poisson manifold. It therefore implies a degeneracy of
the Poisson tensor which cannot be inverted to obtain a symplectic form. The dynamics are nevertheless determined uniquely because Hamilton's equations, used in what follows for evolution as well as gauge transformations, only require a Poisson bracket. 

The effective Hamiltonian $H_{\rm eff}=\langle\hat{H}\rangle$ used in
(\ref{Poisson}) can be interpreted as a function of the moments obtained from
the state that appears in the expectation value.  It can be computed
explicitly to order $N/2$ in $\hbar$, for any integer $N$, by applying a
Taylor expansion to $\langle\hat{H}\rangle$ around any fixed pair of basic
expectation values:
\begin{eqnarray}
 H_{\rm eff} &=& \langle H(\hat{q},\hat{p})\rangle= \langle
 H(q+(\hat{q}-q),p+(\hat{p}-p)\rangle\label{Heffgen} \\
 &=& H(q,p)+\sum_{n+m=2}^{N} \frac{1}{n!m!}
 \frac{\partial^{n+m}H(q,p)}{\partial q^n\partial p^m}
 \Delta(q^np^m)\,. \nonumber
\end{eqnarray}
(Here, we assume that the Hamilton operator is Weyl ordered. For a Hamiltonian
polynomial in $q$ and $p$, the series always truncates at a finite order. It
merely rewrites bare moments $\langle\hat{q}^n\hat{p}^m\rangle$ in terms of
central moments $\Delta(q^np^m)$. These are centered around basic expectation values,
according to (\ref{Delta}). For non-polynomial Hamiltonians, the series
is in general asymptotic.)

Written as a phase-space function, the Hamiltonian (\ref{Heffgen}) generates
equations of motion. This is accomplished by coupling basic expectation values and moments, such as $\langle\hat{q}\rangle$ and $\langle\hat{p}\rangle$, by applying Hamilton's equations with the Poisson bracket (\ref{Poisson}). However, while it can easily be seen that $\{\langle\hat{q}\rangle,\langle\hat{p}\rangle\}=1$ is of canonical form, the moments are not canonical variables. For instance, $\{\Delta(q^2),\Delta(p^2)\}=4\Delta(qp)$. More generally, second-order
moments of $M$ classical degrees of freedom have brackets equivalent to the
Lie algebra ${\rm sp}(2M,{\mathbb R})$ \cite{Bosonize,EffPotRealize}, while
higher-order moments have brackets quadratic in moments
\cite{EffAc,HigherMoments}. 

It is therefore convenient to apply a transformation from moments to canonical
coordinates. Such a transformation always exists locally, according to the
Darboux theorem \cite{Arnold} or its extension to Poisson manifolds
\cite{Weinstein}. To second order for a single classical degree of freedom,
canonical coordinates for the moments $\Delta(q^2)$, $\Delta(qp)$ and
$\Delta(p^2)$ are given by $(s,p_s)$ such that
\cite{VariationalEffAc,EnvQuantumChaos,QHDTunneling}
\begin{equation} \label{sps}
 \Delta(q^2) = s^2 \quad,\quad \Delta(qp)=sp_s \quad,\quad \Delta(p^2)=
 p_s^2+\frac{U}{s^2}
\end{equation}
with a Casimir function $U$, restricted by Heisenberg's uncertainty relation to
obey  the inequality $U\geq\hbar^2/4$.
As a Casimir function, $U$ has vanishing Poisson brackets with any other phase-space function that depends only on basic expectation values and second-order moments. In particular, its Poisson bracket with the Hamiltonian vanishes in a second-order truncation, which means that $U$ is conserved to this order. In quantum mechanics, the phase-space function $U$ is reduced to a constant on any given solution which determines how close the evolving state is to saturating the uncertainty relation. One of the more technical aims of the present paper will be to explore the role of $U$ in a field theory, where it may be a function of spatial coordinates.

Inserting the canonical form (\ref{sps}) of moments in the expansion
(\ref{Heffgen}) for $N=2$, assuming a classical-mechanics Hamiltonian with
generic potential $V(q)$, we obtain
\begin{equation}
 H_{\rm eff}= \frac{p^2}{2m}+ \frac{p_s^2}{2m}+ \frac{U}{2ms^2}+ V(q)+
 \frac{1}{2} V''(q)s^2\,.
\end{equation}
The last three terms together form the effective potential
\begin{equation} \label{Veff}
 V_{\rm eff}(q,s)= \frac{U}{2ms^2}+ V(q)+
 \frac{1}{2} V''(q)s^2\,.
\end{equation} 
The independent quantum degree of freedom $s$ describes quantum
corrections by two terms in the effective potential: The first term, $U/(2ms^2)$, originates in the kinetic energy or momentum fluctuations. In the effective picture, its $U/s^2$-form (where $U$ is strictly positive) prevents position fluctuations $s$ from reaching zero. The other term, $\frac{1}{2}V''(q)s^2$, may be positive or negative depending on the classical potential. It is positive around local minima, where it raises the ground-state energy by a term analogous to zero-point fluctuations. The term is negative around local maxima, which would be relevant in quasiclassical descriptions of tunneling phenomena.

The description is non-adiabatic because no assumption has been
made about the rate of change of $s$ compared with $q$. If, by contrast, one
assumes that $s$ changes slowly and merely tracks its $q$-dependent minimum
\begin{equation}
 s_{\rm min}(q)= \sqrt[4]{\frac{U(x)}{mV''(q)}}
\end{equation}
of a ground state in the potential (\ref{Veff}), one obtains a $q$-dependent effective potential
\begin{equation} \label{Vlowenergy}
 V_{\rm low-energy}(q)  = V(q)+ \sqrt{\frac{U(x)V''(q)}{m}}\,.
\end{equation}
This quasiclassical result equals the standard low-energy effective potential for the minimum value
$U=\hbar^2/4$ \cite{EffAc,EffAcQM}. For the harmonic oscillator, for instance,
$V''(q)=m\omega^2$ implies the correct zero-point energy
$\frac{1}{2}\hbar\omega$. A higher-order adiabatic approximation implies
higher-derivative corrections to the classical equations of motion
\cite{HigherTime}. An adiabatic approximation to all orders would imply a
non-local theory with time derivatives of arbitrarily high orders. Such a non-local theory, which is often complicated because it cannot be analyzed by solving local partial differential equations, can
more easily be studied by keeping $s$ as an independent field in a
non-adiabatic quasiclassical formulation. (From the point of view of the non-local theory, $s$ would be considered an auxiliary field that makes it possible to write non-local equations in local form. Here, however, $s$ has physical meaning; $s$ represents quantum fluctuations in one of the classical degrees of freedom. The local formulation is therefore more physical than an alternative non-local theory obtained by eliminating $s$ by partially solving equations for it in an adiabatic expansion.)

Along similar lines, a canonical moment description of field theories has been
performed in \cite{CW}, where the analog of (\ref{Vlowenergy}) is the
Coleman--Weinberg potential \cite{ColemanWeinberg}. Here, we apply canonical
moment methods to a field theory motivated by spherically symmetric
gravitational systems. This formluation retains independent quantum degrees of freedom, such
as a field version of $s$. We therefore derive non-adiabatic or non-local
effects of quantum gravity.

\subsection{Effective constraints}

Relativistic systems are subject to constraints, instead of Hamiltonian
evolution with respect to an absolute time. The formalism of effective and
quasiclassical methods therefore has to be generalized to constrained systems,
as done in \cite{EffCons,EffConsRel,EffConsQBR}. The main observation is that
the presence of new quantum degrees of freedom, such as $\Delta(q^2)$, implies
additional constraints compared with the classical theory.

An effective constraint, 
\begin{equation} \label{Ceff}
 C_{\rm eff}=\langle\hat{C}\rangle
\end{equation}
for a constraint operator $\hat{C}$, is defined just like an effective
Hamiltonian. An effective constraint is a function on the phase space of basic
expectation values and moments, which can be computed by Taylor expansion as
in (\ref{Heffgen}). The Hamilton's equations generated by $C_{\rm eff}$
correspond to gauge transformations rather than strict evolution. According to
Dirac's quantization procedure for constrained systems, effective constraints
must vanish on physical solutions, $C_{\rm eff}=0$. This is because the constraint
operator $\hat{C}$ annihilates any admissible state upon which it acts. (As a general phase-space function, $C_{\rm eff}$ is obtained for states in the so-called
kinematical Hilbert space of states not necessarily annihilated by $\hat{C}$. In addition, solving the equation $C_{\rm eff}$ implicitly restricts solutions to the
physical Hilbert space of states annihilated by $\hat{C}$.)

Classical constraints, where $C(q,p)$ and $f(q,p)C(q,p)$ as
phase-space functions imply the same gauge flow on the constraint surface, and
have the same solution space as long as $f\not=0$. In contrast, expressions such as
$\langle\hat{C}\rangle$, and $\langle f(\hat{q},\hat{p})\hat{C}\rangle$ in
general, imply independent functions when expressed in terms of basic
expectation values and moments. (For instance, in the simple case of $\hat{C}=\hat{p}$ and $f(\hat{q},\hat{p})$, the constraint $\langle\hat{C}\rangle=\langle\hat{p}\rangle=0$ restricts the expectation value $\langle\hat{p}\rangle$, while $\langle f(\hat{q},\hat{p})\hat{C}\rangle=\langle\hat{p}^2\rangle=0$ then requires zero variance as well. In a kinematical state, the expectation value $\langle\hat{p}$ and the variance $\Delta(p^2)$ can be chosen independently, for instance in a standard Gaussian wave function.) Effective descriptions of singly-constrained classical systems are therefore subject to multiple constraints. These constraints are of a number that depends on the order of moments considered. Based on
\cite{EffCons,EffConsRel}, it is convenient to organize higher-order
constraints by powers of the same basic operators used in the moments that
describe a given system. In addition to $C_{\rm eff}=\langle\hat{C}\rangle$,
we have independent constraints 
\begin{equation} \label{Cqp}
 C_{q^np^n}=\left\langle
 \left((\hat{q}-\langle\hat{q}\rangle)^n(\hat{p}-\langle\hat{p}\rangle)^m
 \right)_{\rm Weyl} \hat{C}\right\rangle
\end{equation}
for integer $n$ and $m$ such that $n+m\geq 1$. 

While we symmetrize products of non-commuting $\hat{q}$ and $\hat{p}$, we have
to keep $\hat{C}$ to the right, to make sure that it always acts on the state
used in the expectation value. In general, higher-order effective constraints
therefore take complex values. Solving them for moments then results in
complex values. This indicates that the inner product used on the kinematical
Hilbert space which defines effective constraint functions is adjusted when a
physical Hilbert space is introduced for the solution space. In an effective
constrained system, the transition from a kinematical to a physical Hilbert space, which can be very
complicated in generic quantum systems and is in general uncontrolled, is
implicitly performed by simply imposing reality conditions for combinations of
moments that solve the constraints. The consistency of this approach has been
demonstrated in several examples
\cite{EffCons,EffConsRel,EffConsComp,EffTime,EffTimeLong,EffTimeCosmo,TwoTimes,EffConsPower}.

Because $\langle\hat{O}-\langle\hat{O}\rangle\rangle=0$ for any operator
$\hat{O}$, all terms in higher-order constraints (\ref{Cqp}) contain at least
one moment factor. Therefore, they can be considered as constraints on the
moments, supplementing the effective constraint (\ref{Ceff}) which restricts
basic expectation values, subject to quantum corrections depending on
moments. Since moments up to a given order in general form a Poisson manifold
that is not symplectic, applying the usual constraint formalism requires a
generalization to Poisson manifolds as given in \cite{brackets}. In
particular, it is possible for a number $N$ of first-class constraints (that
is, $C_i$ with $i=1,\ldots,N$ such that all Poisson brackets
$\{C_i,C_j\}\approx 0$ vanish on the solution space of the constraints $C_i$)
to generate gauge flows that span a hypersurface of dimension less than $N$. 

The formalism of effective constraints has a straightforward generalization to
systems with more than one classical constraint. If the classical constraints
are first class, the corresponding effective and higher-order constraints are
then guaranteed to be first class as well. A new feature arises in constrained
systems with structure functions, as in general relativity. If there is a
first-class quantization with constraint operators $\hat{C}_i$ such that
$[\hat{C}_i,\hat{C}_j]=i\hbar \sum_k\hat{f}_{ij}^k \hat{C}_k$ with
operator-valued coefficients $\hat{f}_{ij}^k$, effective constraints have the
Poisson-bracket relations \cite{EffConsQBR}
\begin{equation} \label{CC}
 \{C_{i,{\rm eff}},C_{j,{\rm eff}}\}= \sum_k \langle \hat{f}_{ij}^k
 \hat{C}_k\rangle= \sum_k f_{ij}^k{}_{\rm eff} C_{k,{\rm eff}} +\cdots
\end{equation}
where $f_{ij}^k{}_{\rm eff}$ are effective structure functions obtained from
$\langle \hat{f}_{ij}^k\rangle$, and the dots indicate neglected higher-order
constraints. For systems with structure functions, the basic effective
constraints (\ref{Ceff}) and higher-order constraints (\ref{Cqp}) are
therefore coupled in the constraint algebra, forming an enlarged system of
underlying gauge symmetries.

It is an interesting question whether such an enlarged system in models of
gravity can be interpreted as an extended space-time structure. Here, we will
not address this question in complete generality because we will restrict our
attention to static solutions. However, our constraints will have higher-order
corrections, allowing us a glimpse on what moment-based extended space-time
structures might entail. In our technical analysis, we will combine the
formalism of effective constraints with a field-theory version of the
canonical variables (\ref{sps}) for moments, restricted to spherical
symmetry. The metric components that determine the fields of spherically
symmetric gravity will be complemented by an additional canonical field,
$\phi_3$, representing quantum fluctuations of $\phi_2$.

\section{Space-time in quasiclassical form}
\label{s:SpaceTime}

In a classical canonical formulation of general relativity, the line element
of spherically symmetric space-times is defined by
\begin{eqnarray}
  {\rm d}s^2&=&
   -N(t,x)^2{\rm d}t^2+ q_{xx}(t,x) \left({\rm d}x+M(t,x) {\rm
                d}t\right)^2\nonumber\\
  &&+ q_{\varphi\varphi}(t,x) {\rm d}\Omega^2 
\end{eqnarray}
with the lapse function $N$, the radial component $M$ of the shift vector,
and two independent spatial metric components, $q_{xx}$ and
$q_{\varphi\varphi}$. 

The definition of a line element entails that it implies coordinate invariant geometrical statements such as distances, areas or volumes as well as physically important concepts such as geodesics or horizons. A geometry described by a line element can therefore be evaluated with any choice of coordinates, or any conditions slicing space-time into spatial hypersurfaces. However, individual metric components such as $N$ or $q_{xx}$ are not invariant and must transform in a specific way under coordinate changes for the line element to be invariant. Classically, this consistency condition is described by the tensor-transformation law for the space-time metric. But it is not clear that quantization (even in a quasiclassical form, which avoids operators but amends terms---such as $N$ and $q_{xx}$---by quantum corrections $\delta N$ and $\delta q_{xx}$) can maintain this condition.

We will use a canonical approach, in which the space-time metric is replaced by time-dependent families of fields (for $q_{xx}(t)$ and $q_{\varphi\varphi}(t)$, as well as their momenta). We will do this such that fixing the value of $t$ is classically equivalent to fixing a constant-$t$ hypersurface in space-time. The lapse function $N$ and shift vector $M$ then appear as coefficients in evolution equations for these fields. These evolution equations are obtained as Hamilton's equations generated by a phase-space function, which can be written in the form $H[N]+D[M]$ with the Hamiltonian constraint $H$ and the diffeomorphism constraint $D$. Since changes of hypersurfaces are gauge transformations, their generators $H$ and $D$ are constrained to vanish. Several consistency conditions then immediately arise, because the constraints $H=0$ and $D=0$ must hold at all times. Therefore, they must be preserved by Hamiltonian evolution generated by $H[N]+D[M]$, and the combination of two slicing changes must be another slicing change. In technical terms, the constraints must therefore be first-class. They must also have Poisson brackets suitable for the geometrical form of hypersurface deformations in space-time. Since it is difficult to evaluate these conditions for quantum-corrected constraints, we will do so here only for a specific class of gauge transformations that preserve the static nature of solutions. We will therefore check that the Poisson brackets of constraints have the correct form, only in the case of vanishing momenta. What we will now refer to as consistency conditions has the following ingredients:
\begin{itemize}
    \item There is a quasiclassical set of constraints, of the form $H+\delta H$ and $D+\delta D$, where $H$ and $D$ are the classical expressions. Additionally, $\delta H$ and $\delta D$ depend on quantum fluctuations, in a specific way derived from the classical constraints following (\ref{Heffgen}).
    \item The constraint brackets remain first class, and of hypersurface-deformation form, when restricted to the phase-space submanifold of vanishing momenta. This value, $\{(H+\delta H)[N_1],(H+\delta H)[N_2]\}$, is proportional to the diffeomorphism constraint, and therefore vanishes when restricted to the submanifold of vanishing momenta. We set the momenta equal to zero, only after evaluating the Poisson bracket, which therefore is not trivially zero.
    \item We will be able to go slightly beyond the preceding condition by comparing momentum-dependent terms in the Poisson bracket $\{(H+\delta H)[N_1],(H+\delta H)[N_2]\}$ with terms expected from the classical structure function resulting from this bracket. Some terms are as expected, but others are not. This observation highlights the necessity of vanishing momenta at the current stage of developments for quasiclassical constraints.
    \item In practical terms, we will explicitly demonstrate that all the constraint and evolution equations of the quasiclassical system have mutually consistent static solutions with the desired classical limit. 
\end{itemize}
We will now recall detailed definitions of the phase-space variables and properties of the constraints.

\subsection{Variables and constraints}

At any $x$, the phase space of metric components has a boundary given by the
inequality $\det q=q_{xx}q_{\varphi\varphi}^2>0$. A canonical quantization of
these variables therefore requires some care \cite{AffineQG,AffineQG2}. Here,
we avoid this issue by using a triad  formulation, with two components $E^x$
and $E^{\varphi}$ of a (densitized) triad at each $x$, related to the metric
components by the canonical transformation
\begin{equation} \label{qE}
 q_{xx} = \frac{(E^{\varphi})^2}{|E^x|} \quad,\quad q_{\varphi\varphi} = |E^x|\,.
\end{equation}
(These components of a spherically symmetric metric are derived from the general relationship $q^{ab}=E^a_iE^{bi}/|\det(E^c_j)|$ between the inverse spatial metric and a densitized triad $E^a_i$.) 
In our explicit calculations, we will assume $E^x>0$, corresponding to a
right-handed triad. But in general, $E^x$, unlike the metric components, may
take negative values for a left-handed triad thanks to absolute values in
(\ref{qE}), and the sign of $E^{\varphi}$ does not matter thanks to the
quadratic appearance in (\ref{qE}). In a triad formulation, it is therefore
possible to apply standard canonical quantization of a phase space without
boundaries.  According to the appearance of $E^x$ and $E^{\varphi}$ in the spatial metric, the former (times $4\pi$) represents the areas of 2-spheres at a constant radial coordinate $x$, while the latter determines the radial distance.

Momenta of the triad fields are classically given by the components of extrinsic
curvature, such that we have basic Poisson brackets
\cite{SymmRed,SphSymm,SphSymmHam} 
\begin{eqnarray} \label{KE}
 \{K_x(x),E^x(y)\} &=& 2G\delta(x,y) \quad\mbox{and}\nonumber\\
 \{K_{\varphi}(x),E^{\varphi}(y)\} &=& G\delta(x,y)
\end{eqnarray}
with Newton's constant $G$.  (There is no factor of two in the second equation because the angular direction represents two degrees of freedom on a 2-sphere that are strictly related by spherical symmetry.) The relationship between $K_x$ and $K_{\varphi}$ and derivatives of the triad components follows from equations of motion of the classical theory. Classically, $K_{\varphi}$ is proportional to the change in time of $E^x$ or of 2-sphere areas, while $K_x$ determines the change in time of the radial distance. These relationships, in general, may be modified by quantum effects introduced in the canonical dynamics. 

Specific equations of motion are generated by a combination of constraints,
the Hamiltonian constraint $H[N]$, corresponding to conservation of energy, and the diffeomorphism constraint $D[M]$, corresponding to conservation of momentum, such that $\dot{f}=\{f,H[N]+D[M]\}$ for any phase-space function $f$. The dot
refers to a time derivative in the direction of an evolution vector field
$t^a=Nn^a+M e^a$ determined by lapse and shift \cite{ADM}, where $n^a$ is the
future-pointing unit normal to a space-like foliation and $e^a$ a unit vector
tangential to the foliation. In classical spherically symmetric gravity, the
constraints as phase-space functions take the form
\begin{eqnarray}
H[N] &=&-\frac{1}{G} \int {\rm d} x\, N \left( \frac{E^{\phi}}{2 \sqrt{E^x}}
         K_{\varphi}^2 +  \sqrt{E^{x}}K_{\varphi}  K_x\right.\\
  &&\hspace{2cm}+ \frac{E^{\phi}}{2
    \sqrt{E^x}}  - \frac{((E^x)')^2}{8 \sqrt{E^{x}} E^{\phi}} \nonumber\\ 
  & & \hspace{2cm} \left. + \frac{\sqrt{E^x} (E^x)' (E^{\phi})'}{2
      (E^{\phi})^2} - \frac{\sqrt{E^x}(E^x)''}{2E^{\phi}}\right)\nonumber
\end{eqnarray}
and
\begin{equation}
D[M] = \frac{1}{2G} \int {\rm d}x\, M \left(2K_\varphi^\prime\, E^\varphi - K_x
  (E^x)^\prime \right)\,.
\end{equation}
They form a first-class system with brackets 
\begin{eqnarray}
 \{D[M_1],D[M_2]\} &=& D[M_1M_2'-M_2 M_1'] \label{DD} \\
 \{H[N],D[M]\} &=& -H[MN'] \label{HD}
 \end{eqnarray}
 and
 \begin{eqnarray}
 &&\{H[N_1],H[N_2]\}\nonumber\\
 &=& - D[E^x(E^{\varphi})^{-2}(N_1N_2'-N_2 N_1')] \label{HH}
\end{eqnarray}
corresponding to deformations of spacelike hypersurfaces in classical space-times with spherical symmetry. The structure function $E^x/(E^{\varphi})^2$ in the last equation is the only component of the inverse spatial metric that contributes if spherical symmetry is imposed.

Any $(1+1)$-dimensional triad theory subject to brackets (\ref{DD}),
(\ref{HD}) and (\ref{HH}) is generally covariant \cite{MidiClass}, in the sense that solutions of the theory are subject to gauge transformations equivalent to space-time coordinate transformations. The general form of the brackets should therefore be maintained by quantum corrections. More
generally, it may be possible that quantum corrections preserve the
first-class nature of the two constraints, $H[N]$ and $D[M]$, but with
modified brackets. In particular, as in (\ref{CC}) the phase-space function
$E^x (E^{\varphi})^{-2}$ in (\ref{HH}) may be quantum corrected if $E^x$ and
$E^{\varphi}$ are quantized. Such a theory would still be consistent, but it
may not describe space-time with Riemannian geometry. It would rather describe
a quantum version of space-time with a structure that depends on the detailed
modification of the coefficient $E^x (E^{\varphi})^{-2}$, or on higher-order
versions (\ref{Cqp}) of the gravitational constraints. 

Quantum corrections considered in the present paper will not present a clear modification of the structure function, but information about this possibility is limited by the restriction to static configurations that we will make for tractable equations. Further analysis of non-static solutions will be necessary before statements about the quasiclassical structure of space-time can be made. Nevertheless, within the setting to be developed here, it is possible to study implications of quantum effects on specific static solutions.
To this end, we initiate and apply here a canonical
description of quasiclassical quantum field theory. The canonical nature makes
it possible to extend the Poisson brackets used in (\ref{DD}) and (\ref{HH})
to constraints amended by quantum corrections. Consistent orderings of
constraint operators are known in spherically symmetric quantum gravity
\cite{SphKl1,SphSymmOp}, which guarantees that closed effective constraint
brackets of the form (\ref{CC}) exist. The required equations have been
derived explicitly in \cite{SphSymmMomentsMasters}, where consistency was
confirmed independently. The quasiclassical nature means that we will be able
to include key features such as quantum fluctuations or uncertainty relations,
in our analysis. (In addition, factor ordering choices matter in quantum
constraints, which in our context imply certain imaginary contributions to
effective constraints that we will not consider in detail here.) 

It will also turn out to be important that the methods we use, which are
generalized versions of what has been known for some time in quantum chemistry
\cite{QHDTunneling}, are non-adiabatic. In our context, this non-adiabaticity
means that we will not be required to express quantum corrections in the form
of a derivative expansion, as implicitly done by common methods of quantum
field theory such as low-energy potentials or Feynman expansions. Quantum
corrections are rather expressed in terms of independent degrees of freedom
that physically correspond to fluctuations or higher moments of a state.

\subsection{Canonical fields} 

Before we implement fluctuation variables, we transform our current fields to
strictly canonical form, removing a factor of two in (\ref{KE}). (From now on,
we choose units such that $2G=1$.) Also renaming the fields, this
transformation is accomplished by introducing
\begin{equation}
 \phi_1=E^x\quad,\quad p_1=-K_x \quad,\quad \phi_2=2E^{\varphi} \quad,\quad
 p_2=-K_{\varphi}\,.
\end{equation}
In this notation, $\phi_1$ and $\phi_2$ therefore represent the metric components with momenta $p_1$ and $p_2$. Classically, the momentum fields have the usual interpretation as extrinsic curvature, but this relationship will be modified by quantum corrections. In these variables, the Hamiltonian constraint takes the form
\begin{equation} \label{Hclass}
 H[N]=-\int{\rm d}x N(x) \left(\frac{\phi_2p_2^2}{2\sqrt{\phi_1}}+
   2\sqrt{\phi_1} p_1p_2+ \left(1-\left(\frac{\phi_1'}{\phi_2}\right)^2\right)
   \frac{\phi_2}{2\sqrt{\phi_1}}- 2\left(\frac{\phi_1'}{\phi_2}\right)'
   \sqrt{\phi_1}\right)
\end{equation}
while
\begin{equation}
 D[M]= \int{\rm d}xM(x) \left(-\phi_1'p_1+p_2'\phi_2\right)\,.
\end{equation}

The number of independent fields can be reduced by making a gauge choice for
$E^x$ or $\phi_1$ such that $x$ is the usual area radius: $\phi_1=x^2$. The gauge-fixing condition, $g(x)=\phi_1(x)-x^2$ for all $x$, then forms a second-class pair of constraints, together with the diffeomorphism constraint. This is because $\{g(x),D[M]\}= -2M(x)\phi_1(x)\phi_1'(x)\approx -4M(x)x^3\not=0$, unless $x=0$ or $M(x)=0$. Here, $\approx$ indicates that we have used $g(x)=0$ in this step. For second-class constraints, we have to solve both conditions, $D[M]=0$ for all $M$ and $g(x)=0$ for all $x$. We do this while removing the diffeomorphism constraint and fixing its gauge freedom, by using a specific radial coordinate $x$, such that $\phi_1(x)=x^2$. In the static case, $D[M]$ is automatically zero. However, its gauge flow, restricted to the submanifold of zero momenta in phase space, does not identically vanish. This is because it may still change $\phi_1$ and $\phi_2$. This freedom is fixed by imposing the condition $g(x)=0$.

The
remaining flow generated by the Hamiltonian constraint will be time evolution
for a given lapse function $N$. The only fluctuating field will then be
$\phi_2$, for which we introduce an independent quantum degree of freedom
$\phi_3$ as a field version of $s=\Delta q$ as recalled for quantum mechanics in Section~\ref{s:EffHam},  together with a momentum field
$p_3$. Therefore, 
\begin{equation} \label{phip}
 \Delta(\phi_2^2)=\phi_3^2\quad,\quad \Delta(\phi_2p_2)=\phi_3p_3 \quad,\quad
 \Delta(p_2^2)= p_3^2+\frac{U(x)}{\phi_3^2}\,.
\end{equation}
As our notation indicates, the Casimir function $U$, which was a function on phase space but constant along solutions in quasiclassical quantum mechanics, may now be a function of the spatial coordinate $x$, just like the other canonical fields. There are no equations of motion for $U$ because it does not have a momentum field. One of the aims of this paper is to look for additional consistency conditions that may be used to determine $U$ based on $U$-dependent equations of motion for the other fields.

In order to determine how the new fields appear in an effective Hamiltonian,
we need to perform a Taylor expansion of $H[N]$ by $\phi_2$ and $p_2$, which
is rather lengthy. The result is that the effective Hamiltonian constraint is
of the form
\begin{equation}
 \bar{H}[N]= H[N]+H_2[N]
\end{equation}
with the classical $H[N]$ from (\ref{Hclass}) and a correction 
\begin{eqnarray}
 H_2[N] &=& \int{\rm d}x N(x)
 \left(\frac{1}{2}\frac{\partial^2H}{\partial p_2^2}
   \left(p_3^2+\frac{U}{\phi_3^2}\right)+
   \frac{\partial^2H}{\partial\phi_2\partial p_2}\phi_3p_3
+\frac{1}{2}\frac{\partial^2H}{\partial\phi_2^2} \phi_3^2+
   \frac{\partial^2H}{\partial\phi_2\partial\phi_2'}
   \phi_3\phi_3'\right)\nonumber\\ 
&=& -\int{\rm d}xN(x) \left(\frac{\phi_2p_3^2}{2\sqrt{\phi_1}}+
   \frac{\phi_3p_2p_3}{\sqrt{\phi_1}}+
   \left(6\frac{\sqrt{\phi_1}\phi_1'\phi_2'}{\phi_2^4}- \frac{1}{2}
     \frac{(\phi_1')^2}{\sqrt{\phi_1}\phi_2^3}-
     2\frac{\phi_1''\sqrt{\phi_1}}{\phi_2^3}\right) \phi_3^2\right.\nonumber\\
&&\left.-
   4\frac{\sqrt{\phi_1}\phi_1'\phi_3\phi_3'}{\phi_2^3}+\frac{U(x)\phi_2}{2
     \sqrt{\phi_1}\phi_3^2} \right)\,. \label{H2}
\end{eqnarray}
As in (\ref{Heffgen}), the effective Hamiltonian follows from a Taylor expansion, here in terms of $\phi_2(x)$ at any $x$. The leading corrections are expressed in terms of second-order partial derivatives in the first line of the preceding equation, which are evaluated in the next two lines. 

The last term in (\ref{H2}), $\frac{1}{2}U(x) \phi_2 \phi_1^{-1/2}\phi_3^{-2}$, is implied by
(\ref{phip}). Its analog in quantum mechanics has a contribution from
zero-point fluctuations \cite{CW} that would be subtracted out in a quantum
field theory, or be subject to renormalization. (See also the simple example
we gave after (\ref{Vlowenergy}).) For this reason, and because uncertainty relations for operator-valued fields are less clear than those of quantum mechanics, we will not impose a non-zero lower bound on $U(x)$ such as $\hbar^2/4$. We will, however, require that $U(x)$ be positive for all $x$, motivated by its interpretation as a remnant of zero-point fluctuations. The value of $U(x)$ at a given position can then be used as an indication of the strength of quantum effects.

From the perspective of hypersurface deformation generators, the $U$-term in (\ref{H2}) does not contribute to the Poisson bracket of two Hamiltonian constraints because it does not contain any spatial derivatives or momenta. Therefore,
it does not have an effect on the main consistency test performed in this paper, given by closure of the quasiclassical constraints in the static limit. The term will, however, affect our static solutions to be derived below. 

\subsection{Spatial diffeomorphisms}

It is noteworthy that the function $U(x)$, according to its first appearance in (\ref{phip}), should have spatial density weight two so as to be consistent with a density weight one of $\phi_3$ (inherited from $\phi_2$) and density weight zero of $p_3$. This property provides further motivation for allowing $U(x)$ to be a function of $x$, rather than a constant which would be possible for a density only in a specific spatial coordinate choice. Moreover, any lower bound such as $\hbar^2/4$, imposed on a density, would not be respected by transformations of the spatial coordinate, while positivity $U(x)\geq 0$ is compatible with a density weight. 

The density weight of $U(x)$ also implies that the $U$-term in (\ref{H2}) has the correct density weight
one, as expected for any contribution to a spatial integrand. If the density weight were ignored, the
term would have density weight minus one because $\phi_2$ and $\phi_3$ have the same transformation property according to (\ref{phip}), and $\phi_2$ has
density weight one. This unconventional transformation behavior, if it were used, would be analogous to a property studied in
the minisuperspace context \cite{Infrared,MiniSup,Claims}, where it originated in a contribution to the dynamics from infrared modes included in a symmetric model. Spatially homogeneous minisuperspace models do not provide control over the density weight because the spatial dependence of all functions is ignored. The present paper is the first one that studies this phenomenon in a field-theory setting in which density weights can be determined unambiguously. We will see below that the density
weight of $U(x)$ may be ignored consistently if only spatial transformations are considered that are generated by a quantum corrected diffeomorphism constraint equal to the Poisson bracket of two Hamiltonian constraints (including the structure function). This generator is sufficient for formal consistency of the quasiclassical constraints.  However, if one tries to analyze full covariance under all spatial coordinate transformations, which lies outside the scope of the present paper, there may be further subtleties related to spatial transformations in spatially inhomogeneous quantum midisuperspace models.

The effective diffeomorphism constraint does not follow directly from the quantum-mechanics model because its structure is rather different from a
Hamiltonian. However, we may expect that the effective diffeomorphism constraint should be of the form
\begin{equation} \label{Dbar}
 \bar{D}[M]= \int{\rm d}x M(x) (-\phi_1'p_1+p_2'\phi_2+p_3'\phi_3)\,.
\end{equation}
Unlike $\phi_1$, which transforms as a standard scalar field  in the symmetry reduced model, the field $\phi_2$ transforms with density weight one, a property that is inherited by the original appearance of $\phi_2$ in the metric components. The new field $\phi_3$, which represents quantum fluctuations of $\phi_2$, is assigned the same density weight. These considerations explain the different signs and positions of spatial derivatives in the three terms of (\ref{Dbar}).

 The new term, compared with the classical constraint, can be derived from a quantized $p_2'\phi_2$ after applying a
point-splitting procedure: In order to evaluate the expectation value of a product of field operators, defining the effective diffeomorphism constraint, we follow the quantum mechanics example of (\ref{Heffgen}). We first introduce two slightly different positions
for $\hat{p}_2'(x)$ and $\hat{\phi}_2(y)$, such that the prime uniquely refers to a derivative only of $\hat{p}_2$. This holds, even in a product of these two operators or in the quantum covariance $\langle \hat{p}_2'(x)\hat{\phi}_2(y)\rangle_{\rm symm}={\rm d}\langle \hat{p}_2(x)\hat{\phi}_2(y)\rangle_{\rm symm}/{\rm d}x$. Taking the limit $x\to y$ after moving the derivative out of the expectation value, we obtain
 $\langle \hat{p}_2'(x)\hat{\phi}_2(x)\rangle_{\rm symm}= \lim_{y\to x} {\rm d} \langle \hat{p}_2(x)\hat{\phi}_2(y)\rangle_{\rm symm}/{\rm d}x$, without any ambiguity as to which operator the derivative is acting on. Continuing with this equation, we have
\begin{eqnarray}
  \langle \hat{p}_2'(x)\hat{\phi}_2(x)\rangle_{\rm symm}&=& \lim_{y\to x} \frac{{\rm
      d}}{{\rm d}x} \langle \hat{p}_2(x)\hat{\phi}_2(y)\rangle_{\rm symm}=
  \lim_{y\to 
    x}\frac{{\rm 
      d}}{{\rm d}x} \left(\langle\hat{p}_2(x)\rangle
    \langle\hat{\phi}_2(y)\rangle+  \Delta(p_2(x)\phi_2(y))\right)\nonumber\\
  &=& \lim_{y\to x}\frac{{\rm
      d}}{{\rm d}x} \left(\langle\hat{p}_2(x)\rangle
    \langle\hat{\phi}_2(y)\rangle+  p_3(x)\phi_3(y)\right)
  = p_2'\phi_2+p_3'\phi_3\,.
\end{eqnarray}
(We may assume the symmetric ordering of $\hat{p}_2'$ and $\hat{\phi}_2$
because reordering terms of the quadratic expression would merely be introduce
constants.)  This form of the diffeomorphism constraint is also consistent
with the transformation behavior of $\phi_3$ which, like $\phi_2$, should be a
scalar density, as already observed in (\ref{Dbar}).

A schematic operator version of the diffeomorphism constraint can also be used
to determine which higher-order constraints should contribute to the effective
constraint brackets, as in (\ref{CC}). The classical bracket (\ref{HH}), after
fixing $\phi_1=E^x=x^2$ to be non-dynamical, shows that the two expectation
values $-4\langle \hat{\phi}_2^{-2} \phi_1 \phi_1' p_1\rangle$ and
$4\langle\phi_1\hat{\phi}_2^{-1} \hat{p}_2'\rangle$ will be relevant, which we
should expand by moments of $\phi_2$ and $p_2$. Ignoring ordering questions
for now, we therefore expect the replacements
\begin{eqnarray}
 -4\frac{\phi_1\phi_1'p_1}{\phi_2^2} &\rightarrow&
 -4\left\langle\frac{\phi_1\phi_1'p_1}{(\phi_2+
                                                   \widehat{\Delta\phi_2})^2}\right\rangle\nonumber\\
  &\sim&
-4\frac{\phi_1\phi_1'p_1}{\phi_2^2}  -12
                                                   \frac{\phi_1\phi_1'\phi_3^2p_1}{\phi_2^2} \label{Expand1}
\end{eqnarray}
and
\begin{eqnarray}
4\frac{\phi_1p_2'}{\phi_2} &\rightarrow& 4\left\langle
  \frac{\phi_1(p_2+\widehat{\Delta
      p_2})'}{\phi_2+\widehat{\Delta\phi_2}}\right\rangle\nonumber\\
&\sim& 4\frac{\phi_1p_2'}{\phi_2}+ \frac{\phi_1\phi_3^2p_2'}{\phi_2^3}-
4\frac{\phi_1\phi_3p_3'}{\phi_2^2}  \label{Expand2}
\end{eqnarray}
where $\widehat{\Delta\phi_2}=\hat{\phi}_2-\phi_2$ and $\widehat{\Delta
  p_2}=\hat{p}_2-p_2$. Interestingly, the last term in the preceding equation
cancels out completely with the last term in (\ref{Dbar}) once the latter
equation is evaluated with the structure function according to (\ref{HH}). We
therefore do not expect a term proportional to $p_3'$ in the bracket of two
Hamiltonian constraints, even though it appears in (\ref{Dbar}). 

With this result we can return to the $U$-term in (\ref{H2}), proportional to $U\phi_2/(\sqrt{\phi_1}\phi_3^2)$. Even if the density weight of $U(x)$ is ignored, this term is consistent with gauge transformations generated by a quantum-corrected diffeomorphism constraint that includes the structure function expected for the bracket of two Hamiltonian constraints: Due to the fact that the $p_3$-term is expected to cancel out in this expression, these gauge transformations do not act on the $\phi_3$-dependence of the $U$-term. If this dependence is ignored for the purpose of counting density weights relevant for a Poisson bracket with the diffeomorphism constraint, the remaining dependence on $\phi_2$ provides the expected density weight of one, suitable for an integrand. (If the density weight of $\phi_3$ is included in the count, one has to assign a density weight two to $U(x)$. As already mentioned, this definition is likely necessary if one attempts to extend diffeomorphism to arbitrary shift vectors. If the structure function is not included in the diffeomorphism constraint, the latter depends on $p_3$ and is sensitive to the density weight of $|\phi_3\rangle$.)

As a further test of mutual consistency of the quasiclassical constraints, we now evaluate the bracket of two Hamiltonian constraints in more detail.  The derivation of
$\{\bar{H}[N],\bar{H}[M]\}$ can be split up into smaller calculations using
\begin{eqnarray}
  && \{\bar{H}[N],\bar{H}[M]\}\nonumber\\
  &=& \{H[N],H[M]\}+ \{H[N],H_2[M]\}+
 \{H_2[N],H[M]\}+
 \{H_2[N],H_2[M]\} \nonumber\\
 &=& \{H[N],H[M]\}+ \{H[N],H_2[M]\}- \{H[M],H_2[N]\}+
 \{H_2[N],H_2[M]\} \label{HHsplit}
\end{eqnarray}
based on the antisymmetry of the Poisson bracket. We already have the first
term in (\ref{HHsplit}), so we only need to derive the second term,
$\{H[N],H_2[M]\}$, and the last term, $\{H_2[N],H_2[M]\}$. The third term can
then be obtained from the second term by flipping $N$ and $M$. The last
bracket, $\{H_2[N],H_2[M]\}$, and the combination $\{H[N],H_2[M]\}-
\{H[M],H_2[N]\}$ are antisymmetric in $N$ and $M$. It is therefore sufficient
to consider only terms in which a spatial derivative of $\phi_2$ or $\phi_3$
appears, which after integration by parts then leads to the non-zero
antisymmetric combination $NM'-N'M$. We are interested here in the Poisson brackets of gauge generators, relevant for our consistency conditions, as well as asymptotically flat solutions close to the classical case for large $x$. Therefore, the lapse function is required to drop off to zero at infinity, while the $\phi$-dependent terms in the constraint remain finite. (The fields $\phi_1$ and $\phi_2$ asymptotically grow like $x^2$ and $x$, respectively, but the momentum-independent terms in the Hamiltonian constraint contain only ratios or derivatives of these fields with finite limits.) We can then ignore boundary terms when integrating by parts for gauge generators. Lapse functions with non-zero limits at infinity correspond to symmetry generators in the asymptotically flat region, which we do not consider here.

A lengthy calculation produces the result 
\begin{eqnarray}
  && \{\bar{H}[N],\bar{H}[N]\}\nonumber\\
  &=& \int{\rm d}x (NM'-N'M)
 \left(-4\frac{\phi_1\phi_1'}{\phi_2^2}p_1+
   4\frac{\phi_1}{\phi_2}p_2'\right.+\left.12\frac{\phi_1\phi_1'\phi_3^2}{\phi_2^4}p_1- 4
   \frac{\phi_1\phi_3^2}{\phi_2^3} p_2'-
   2\frac{\phi_1'\phi_3}{\phi_2^2}p_3\right)\nonumber\\
   &=&  \int{\rm d}x (NM'-N'M) \frac{4\phi_1}{\phi_2^2}
 \left(-\phi_1'p_1+
   \phi_2p_2'+3\frac{\phi_3^2}{\phi_2^2}\phi_1'p_1- 
   \frac{\phi_3^2}{\phi_2} p_2'-
   \frac{1}{2}\phi_1'\phi_3p_3\right)
\end{eqnarray}
for the Poisson bracket of two Hamiltonian constraints. The first two terms
are the classical diffeomorphism constraint with the correct structure
function, while the next two terms are quantum corrections as expected from
the expansions (\ref{Expand1}) and (\ref{Expand2}). 
The last term does not correspond to a contribution in the diffeomorphism
constraint. It can be seen as a consequence of our reduction, which includes
quantum corrections only of $\phi_2$ but not of $\phi_1$. In particular, a
complete effective constraint would include moments such as
$\Delta(\phi_2\phi_1')$ as well as $\Delta(p_1p_2)$ with a bracket that can
contribute to the $\phi_3p_3$-term we obtained here. As shown by the
consistency check in \cite{SphSymmMomentsMasters}, all such contributions
indeed cancel out in the complete effective system, while they do not
completely cancel out in our reduction. The left-over contribution here
re-introduces a $p_3$-dependence that generates non-trivial transformations on
the $U$-term in (\ref{H2}). Our system is therefore not fully consistent if generic spherically symmetric configurations are considered, but
it may be used for static solutions for which the last term in
$\{\bar{H}[N],\bar{H}[N]\}$ vanishes. The solutions we obtain are also
reliable as consistent configurations in the complete system in which all
cross-correlations between $\phi_1$ and $\phi_2$ vanish.

\subsection{Higher-order constraint}

In addition to $\bar{H}[N]$ and $D[M]$, there is one higher-order constraint
of the form (\ref{Cqp}) that is relevant for static solutions at second order
in moments:
\begin{equation}
 H_{\phi_2}[L]=\langle(\hat{\phi}_2-\langle\hat{\phi}_2\rangle)
 \hat{H}[L]\rangle\,.  
\end{equation}
This constraint does not directly contribute to the brackets of hypersurface deformations, but it provides additional restrictions on the fields that are implied by imposing the quantum constraint.
For a derivation of $H_{\phi_2}[L]$ in terms of moments, we need a Taylor
expansion of $\hat{H}[L]$ to first order in
$\hat{\phi}_2-\langle\hat{\phi}_2\rangle$ and
$\hat{p}_2-\langle\hat{p}_2\rangle$. These terms, together with the factor of $\langle\hat{\phi}_2-\langle\hat{\phi}_2\rangle\rangle$ included in the definition of $H_{\phi_2}[L]$, then produce second-order moments. Considering the fact that $\hat{H}$ locally depends $\phi_2$ as well as on $\hat{\phi}_2'$, we write
\begin{eqnarray}
 H_{\phi_2}[L] &=& \int{\rm d}x L(x) \left(\frac{\partial H}{\partial\phi_2}
   \phi_3^2+ \frac{\partial H}{\partial \phi_2'} \phi_3\phi_3'+ \frac{\partial
     H}{\partial p_2} \phi_3p_3\right)\nonumber\\
&=&-\int{\rm d}x L(x)\Biggl( \left(\frac{p_2^2}{2\sqrt{\phi_1}}+
  \frac{1}{2\sqrt{\phi_1}}+ 
  \frac{(\phi_1')^2+4\phi_1\phi_1''}{2\sqrt{\phi_1}\phi_2^2}
  -4\frac{\sqrt{\phi_1}\phi_1'\phi_2'}{\phi_2^3}\right) \phi_3^2\nonumber\\
&& +\frac{2\sqrt{\phi_1}\phi_1'}{\phi_2^2} \phi_3\phi_3'+
\left(\frac{\phi_2p_2}{\sqrt{\phi_1}}+ 2\sqrt{\phi_1}p_1\right) \label{Hphi2}
\phi_3p_3\Biggr) \,.
\end{eqnarray}
In the first line, the common factor of $\phi_2$ in all three terms is implied by the explicit factor of $\langle\hat{\phi}_2-\langle\hat{\phi}_2\rangle\rangle$ in the definition of $H_{\phi_2}[L]$. The remaining factors of $\phi_3$, $\phi_3'$ and $p_3$, respectively, are correspond to first-order terms in a Taylor expansion of $H$.

The presence of higher-order constraints implies that evolution is not
uniquely determined by the classical pair of two functions, lapse and shift, but also
requires the specification of additional functions such as $L$. The latter
determine the direction of a time evolution vector field in moment or state space. The
general form of evolution equations with moment terms is therefore given by
\begin{equation}
 \dot{f}= \{f,\bar{H}[N]+D[M]+H_{\phi_2}[L]+\cdots\}
\end{equation}
for any phase-space function $f$, where the dots indicate further higher-order
constraints that would involve higher moments or higher-order versions of the
diffeomorphism constraint. The former do not appear to second order as
considered here, while the latter, just like the classical $D[M]$, is not
included for static solutions. Our evolution equations will therefore be given
by $\dot{f}= \{f,\bar{H}[N]+H_{\phi_2}[L]\}$. In the static case, $N$ is classically determined by the consistency condition that evolution equations be compatible with static behavior. As we will see, the same is true for $L$ if quantum fluctuations are required to be static too.

\subsection{Solutions}

We will derive properties of static solutions in radial gauge, choosing
$\phi_1=x^2$ such that $x$ is the area radius. Since our extended system is
first class according to (\ref{CC}), we are allowed to fix the gauge in order
to determine solutions. All momenta vanish for static solutions, and we are
left with four free functions, $\phi_2$, $\phi_3$, $N$ and $L$. 

\subsubsection{Equations}

The diffeomorphism constraint (along with its higher-order versions) is
identically satisfied for static solutions, and we have fixed its flow. Two
equations of motion,
\begin{eqnarray}
    \dot{\phi}_2 &=& \{\phi_2, \bar{H}[N]+H_{\phi_2}[L] \} = \frac{\delta
      \bar{H}[N]}{\delta p_2}+\frac{\delta H_{\phi_2}[L]}{\delta p_2} \\
 &=& \left(2x p_1+\frac{\phi_2 p_2}{x}+\frac{\phi_3 p_3}{x}\right)N+
 \frac{\phi_3}{x}\left(\phi_3p_2+ \phi_2   p_3\right)L  \nonumber
\end{eqnarray}
and
\begin{eqnarray}
    \dot{\phi}_3 &=& \{\phi_3, \bar{H}[N]+H_{\phi_2}[L] \} = \frac{\delta
      \bar{H}[N]}{\delta p_3}+\frac{\delta H_{\phi_2}[L]}{\delta p_3}\\
 &=&  \frac{1}{x}(\phi_2 p_3+\phi_3 p_2)N +
 \left(\frac{\phi_2p_2}{\sqrt{\phi_1}}+ 2\sqrt{\phi_1}p_1\right) \phi_3L\,, \nonumber
\end{eqnarray}
are identically satisfied in the static case.

The remaining equations are therefore given by two constraints, $\bar{H}[N]=0$
and $H_{\phi_2}[L]=0$, and two equations of motion,
\begin{equation}
 0=\dot{p}_2=\{p_2,\bar{H}[N]+H_{\phi_2}[L]\}= -\frac{\delta \bar{H}[N]}{\delta
   \phi_2}-\frac{\delta H_{\phi_2}[L]}{\delta \phi_2}
   \end{equation}
   and
   \begin{equation}
 0=\dot{p}_3=\{p_3,\bar{H}[N]+H_{\phi_2}[L]\}=-\frac{\delta \bar{H}[N]}{\delta
   \phi_3}-\frac{\delta H_{\phi_2}[L]}{\delta \phi_3}
\end{equation}
in static form. These implement  the correct flow generated by the Hamiltonian
constraint as required for static solutions.  The full equations are rather
lengthy, and will be shown in a more specific form when we start solving them
below. 
With these conditions, we obtain the Hamiltonian constraint
\begin{equation} \label{Hbarx}
 \bar{H}[N]=-\int{\rm d}x N(x) \left(
   \frac{\phi_2}{2x}-\frac{2x}{\phi_2}
   - 4x\left(\frac{x}{\phi_2}\right)'
+   \left(12\frac{x^2\phi_2'}{\phi_2^4}- 
     \frac{6x}{\phi_2^3}\right) \phi_3^2  - 8\frac{x^2\phi_3\phi_3'}{\phi_2^3}+\frac{U\phi_2}{
     2x\phi_3^2}\right)\,,
\end{equation}
the higher-order constraint
\begin{equation}
 H_{\phi_2}[L] =-\int{\rm d}x L(x)\left(\left(
  \frac{1}{2x}+ 
  \frac{6x}{\phi_2^2}
  -8\frac{x^2\phi_2'}{\phi_2^3}\right) \phi_3^2
 +\frac{4x^2}{\phi_2^2} \phi_3\phi_3'\right)
\end{equation}
and the two equations of motion.

These four equations are coupled differential equations for the four free
functions. In order to simplify the solution procedure, we proceed
perturbatively and assume that $\phi_2$ and $N$ are given by their classical
solutions (according to the Schwarzschild line element) plus small corrections
of the order of $\phi_3^2$:
\begin{eqnarray} \label{phi2N}
 \phi_2&=&\phi_2^{(0)}+\delta\phi_2=\frac{2x}{\sqrt{1-\mu/x}} +\delta\phi_2\\
 N&=&N^{(0)}+\delta N=\sqrt{1-\mu/x}+\delta N\nonumber
\end{eqnarray}
where $\phi_2^{(0)}$ and $N^{(0)}$ are obtained from the Schwarzschild line
element. The constant $\mu$ is equal to the mass in our units, having set $2G$ equal to one in order to simplify several numerical factors in the constraints and Poisson brackets. Transforming to the more standard choice where $G$ equals one can easily be achieved by equating $\mu$ to twice the mass. 

The higher-order constraint equation $H_{\phi_2}[L]=0$ then takes
the form
\begin{eqnarray}
  0&=&H_{\phi_2}[L]\\
  &=&-\int{\rm d}x L(x)\phi_3\left(\frac{3\mu}{2x^2}\phi_3+
  \left(1-\frac{\mu}{x}\right) \phi_3'\right)+ O(\phi_3^2\delta\phi_2) \nonumber
\end{eqnarray}
and can be interpreted as a first-order differential equation for
$\phi_3$. Its general solution is
\begin{equation} \label{phi3}
 \phi_3(x)= \frac{C}{(1-\mu/x)^{3/2}}
\end{equation}
with an integration constant $C$. This solution diverges at the horizon, which
is not surprising because this is where our background solutions (\ref{phi2N})
break down in the Schwarzschild coordinate system. At spatial infinity,
$\phi_3$ approaches a constant while $\phi_2$ diverges. Fluctuations are
therefore small at low curvature.

\subsubsection{Metric correction}

Using our solution for $\phi_3$, the constraint (\ref{Hbarx}) implies a
differential equation for $\delta\phi_2$, coupled to $\delta N$.  Only the
classical part of the constraint contributes to the dependence on $\delta
\phi_2$ and $\delta N$ in our perturbative treatment because $H_2[N]$ is  quadratic in the small $\phi_3$, such that any contribution from $\delta\phi_2$ or $\delta N$ would be of higher order. For this contribution, we have
\begin{eqnarray}
 H[N] &=& H[N]|_{\phi_2^{(0)}}
+\int{\rm d}x(N^{(0)}+\delta N)
 \left(\frac{\partial   H}{\partial\phi_2} \delta\phi_2+
          \frac{\partial   H}{\partial\phi_2'} \delta\phi_2'\right)\nonumber\\
  &&
+ \int{\rm d}x N^{(0)} \left(
   \frac{1}{2}\frac{\partial^2H}{\partial\phi_2^2}(\delta\phi_2)^2+
   \frac{\partial^2H}{\partial\phi_2\partial\phi_2'}
   \delta\phi_2\delta\phi_2'\right)
\end{eqnarray}
where all coefficients are evaluated at the classical solution. Therefore,
$H[N]|_{\phi_2^{(0)}}$ is set to zero by definition of $\phi_2^{(0)}$ There is no second-order term in $\delta\phi_2'$ because
the dependence of $H[N]$ on $\phi_2'$ is linear.  In the first line, we can
integrate by parts in the last term. Several resulting contributions then
equal the integral of $\delta\phi_2$ times the classical
\begin{eqnarray} \label{phi2dot}
&& -\dot{p}_2(y)|_{N^{(0)}}= -\{p_2(y),H[N^{(0)}]\}= \frac{\delta H[N^{(0)}]}{\delta\phi_2(y)}\\
 &=& \int{\rm d}xN^{(0)}(x) \left(\frac{\partial H(x)}{\partial \phi_2(y)} \delta(x-y)+ \frac{\partial H(x)} {\partial \phi_2'(y)} \frac{\partial \delta(x-y)}{\partial x}\right)=
 N^{(0)}\frac{\partial H}{\partial\phi_2} -
 \left(N^{(0)}\frac{\partial 
     H}{\partial\phi_2'}\right)' \nonumber
\end{eqnarray}
which vanishes for static background solutions. For the $\delta N$-terms, we
can also integrate by parts,
\begin{eqnarray}
&& \int{\rm d}x\delta N
 \left(\frac{\partial   H}{\partial\phi_2} \delta\phi_2+
\frac{\partial   H}{\partial\phi_2'} \delta\phi_2'\right)= \int{\rm
d}x\frac{\delta N}{N^{(0)}} 
 \left(N^{(0)}\frac{\partial   H}{\partial\phi_2} \delta\phi_2+
N^{(0)}\frac{\partial   H}{\partial\phi_2'} \delta\phi_2'\right) \nonumber\\
&=& \int{\rm
d}x \delta\phi_2 \left(\left(N^{(0)}\frac{\partial   H}{\partial\phi_2}-
\left(N^{(0)}\frac{\partial   H}{\partial\phi_2'}\right)' \right)\delta N-
N^{(0)}\frac{\partial   H}{\partial\phi_2'} \left(\frac{\delta
    N}{N^{(0)}}\right)' \right)
\end{eqnarray}
The first $\delta N$-term in this expression vanishes, again by virtue of
(\ref{phi2dot}), but one term now remains, containing $(\delta N/N^{(0)})'$. Including
this term in the expanded Hamiltonian constraint, we are left with
\begin{equation} \label{Hsecond}
 H[N]= \int{\rm
  d}xN^{(0)} \left(-\frac{\partial H}{\partial \phi_2'}\delta\phi_2
  \left(\frac{\delta 
     N}{N^{(0)}}\right)'+
  \frac{1}{2}\frac{\partial^2H}{\partial\phi_2^2}(\delta\phi_2)^2+
  \frac{\partial^2H}{\partial\phi_2\partial\phi_2'}
  \delta\phi_2\delta\phi_2'\right)\,.
\end{equation}

The expansion of $\bar{H}[N]$ contributes additional terms depending on
$\phi_3$, which by construction of $H_2[N]$ from a Taylor expansion have the
same coefficients as the last two $\delta\phi_2$-terms in (\ref{Hsecond}). All
these terms can be combined to
\begin{eqnarray} \label{Hsecond2}
 \bar{H}[N]&=& \int{\rm
  d}xN^{(0)} \left(-\frac{\partial H}{\partial \phi_2'}\delta\phi_2
  \left(\frac{\delta 
     N}{N^{(0)}}\right)'\right.\\
&&+\left.
  \frac{1}{2}\frac{\partial^2H}{\partial\phi_2^2}
\left((\delta\phi_2)^2+\phi_3^2\right)  +
  \frac{1}{2}\frac{\partial^2H}{\partial\phi_2\partial\phi_2'}
  \left((\delta\phi_2)^2+\phi_3^2\right)'+
  \frac{U(x)\phi_2}{2\sqrt{\phi_1}\phi_3^2}\right)\,. \nonumber
\end{eqnarray}
(The $U$-term should be considered second-order because it is derived from
$\Delta(p_2^2)$ in (\ref{phip}). The function $U(x)$ is therefore of fourth order in the quasiclassical expansion. If moments of a semiclassical or Gaussian state are used, one order in the quasiclassical expansion corresponds to a factor of $\sqrt{\hbar}$.)  Inserting classical solutions, from the background upon which we evaluate the perturbations, in the
coefficients, we obtain
\begin{eqnarray} \label{Hsecondx}
 \bar{H}[N]&=& \int{\rm
  d}x \left(-\left(1-\frac{\mu}{x}\right) \delta\phi_2\delta N'+
  \frac{\mu}{2x^2}\delta\phi_2\delta N
  +\frac{U(x)}{\phi_3^2} 
\right.\\ 
&&+\left. \frac{3}{4x^2}
   \left(1-\frac{\mu}{x}\right) \left(1-\frac{2\mu}{x}\right)
   \left((\delta\phi_2)^2+\phi_3^2\right)- 
   \frac{1}{2x} \left(1-\frac{\mu}{x}\right)^2
   \left((\delta\phi_2)^2+\phi_3^2\right)'\right)\,.\nonumber
\end{eqnarray}

In order to simplify this expression, we can combine it with the a
non-vanishing contribution to the full $\dot{p}_2$ to linear order in
$\delta\phi_2$ and $\delta N$, which will allow us to eliminate $\delta N$
from (\ref{Hsecond2}). Using (\ref{phi2dot}) for the expanded solution, this
linear contribution to $\dot{\phi}_2$ is given by
\begin{eqnarray}
 \dot{p}_2|_{\rm linear} &=& -\frac{\partial H_{\rm
     linear}}{\partial\phi_2}N^{(0)}- \frac{\partial H^{(0)}}{\partial\phi_2}
 \delta N + \left(\frac{\partial H_{\rm
     linear}}{\partial\phi_2'}N^{(0)}+ \frac{\partial H^{(0)}}{\partial\phi_2'}
 \delta N\right)'\nonumber\\
&=& -\left(\frac{\partial^2H}{\partial\phi_2^2} 
   \delta\phi_2 + \frac{\partial^2H}{\partial\phi_2'\partial\phi_2} 
   \delta\phi_2'- \left(\frac{\partial^2H}{\partial\phi_2'\partial\phi_2} 
   \delta\phi_2\right)'\right)N^{(0)}\nonumber\\
&&+ \frac{\partial^2H}{\partial\phi_2'\partial\phi_2} 
   \delta\phi_2 N^{(0)}{}' -\left(\frac{\partial H}{\partial\phi_2}-
     \left(\frac{\partial H}{\partial \phi_2'}\right)'\right) \delta N+
   \frac{\partial H}{\partial\phi_2'} \delta N'\nonumber\\
&=& -\left(\frac{\partial^2H}{\partial\phi_2^2} 
   - \left(\frac{\partial^2H}{\partial\phi_2'\partial\phi_2}
   \right)' \right)
   \delta\phi_2 N^{(0)}\nonumber\\
&&+ \frac{\partial^2H}{\partial\phi_2'\partial\phi_2} 
   \delta\phi_2 N^{(0)}{}' -\left(\frac{\partial H}{\partial\phi_2}-
     \left(\frac{\partial H}{\partial \phi_2'}\right)'\right) \delta N+
   \frac{\partial H}{\partial\phi_2'} \delta N'\,.
 \end{eqnarray}
Upon inserting background solutions in the
coefficients, the $\delta N$-terms in 
\begin{equation} \label{p2linear}
 \dot{p}_2|_{\rm linear} = -\frac{1}{2x^2}\left(1-\frac{\mu}{x}\right)
 \delta\phi_2- \frac{\mu}{2x^2} \delta N+ \left(1-\frac{\mu}{x}\right)\delta N'=0
\end{equation}
are of the same form as those of (\ref{Hsecond2}) and can therefore be eliminated from this equation.
The simplified second-order constraint,
\begin{eqnarray}
 \bar{H}[N]&=& \int{\rm
  d}x\left(- \frac{1}{2x^2}\left(1-\frac{\mu}{x}\right)
  (\delta\phi_2)^2+\frac{U(x)}{\phi_3^2}  \right.\\
&&+\left. \frac{3}{4x^2}
   \left(1-\frac{\mu}{x}\right) \left(1-\frac{2\mu}{x}\right)
   \left((\delta\phi_2)^2+\phi_3^2\right)- 
   \frac{1}{2x} \left(1-\frac{\mu}{x}\right)^2
   \left((\delta\phi_2)^2+\phi_3^2\right)'\right)\,,\nonumber
\end{eqnarray}
provides a differential equation for $\delta\phi_2$ if we use the
known solution (\ref{phi3}) for $\phi_3$. Keeping some of the $\phi_3^2$-terms
for now, we write this differential equation as an inhomogeneous one for
$(\delta\phi_2)^2+\phi_3^2$:
\begin{eqnarray}
&& \frac{1}{4x^2} \left(1-\frac{\mu}{x}\right) \left(1-\frac{6\mu}{x}\right)
 \left((\delta\phi_2)^2+\phi_3^2\right)- \frac{1}{2x}
 \left(1-\frac{\mu}{x}\right)^2 \left((\delta\phi_2)^2+\phi_3^2\right)'
 \nonumber\\
&=&
 -\frac{C^2}{2x^2(1-\mu/x)^2}-\frac{U(x)}{C^2}\left(1-\frac{\mu}{x}\right)^3\,.
\end{eqnarray}
The corresponding homogeneous equation can easily be solved for
\begin{equation}\label{phiD}
 (\delta\phi_2)^2+\phi_3^2 = \frac{D\sqrt{x}}{(1-\mu/x)^{5/2}}\,,
\end{equation}
which then implies the differential equation
\begin{equation}
 D'=\frac{C^2}{(x-\mu)^{3/2}}+\frac{2U(x)}{C^2}
 \frac{(x-\mu)^{7/2}}{x^3} 
\end{equation}
for solution of the inhomogeneous equation of the form (\ref{phiD}) with
$x$-dependent $D$. Solving this equation, we obtain
\begin{equation}
 (\delta\phi_2)^2+\phi_3^2= \frac{E\sqrt{x}}{(1-\mu/x)^{5/2}}-
 \frac{2C^2}{(1-\mu/x)^3}+ \frac{2\sqrt{x}}{C^2(1-\mu/x)^{5/2}} \int U(x)
 \left(1-\frac{\mu}{x}\right)^{7/2}\sqrt{x} {\rm d}x
\end{equation}
with a new integration constant $E$, or
\begin{equation} \label{phi2}
 \delta\phi_2 = \sqrt{\frac{E\sqrt{x}}{(1-\mu/x)^{5/2}}-
   \frac{3C^2}{(1-\mu/x)^3} + \frac{2\sqrt{x}}{C^2(1-\mu/x)^{5/2}} \int U(x)
 \left(1-\frac{\mu}{x}\right)^{7/2}\sqrt{x} {\rm d}x }\,.
\end{equation}
(For constant $U$, there is a closed-form logarithmic expression for $\int (1-\mu/x)^{7/2}\sqrt{x} {\rm d}x$, but
it is lengthy.)

Notice that the second term dominates near the horizon, where it is negative. The perturbative solution therefore breaks down before the horizon is reached, where $\phi_3$ is large but still finite. For $x\gg \mu$, the dominant behavior
of $\phi_2(x)$ is determined by the last term in (\ref{phi2}), which, for an asymptotically constant $U(x)$, behaves like $Ux^2$ (the integral can then be approximated as $\int x^{1/2}{\rm d}x=\frac{2}{3}x^{3/2}$). In this case, therefore, $\delta\phi_2\sim \sqrt{U}x$ grows with $x$, but so does the classical solution $\phi_2{(0)}$. Since $\phi_2^{(0)}\sim x$ for $x\gg \mu$, the ratio $(\delta\phi_2)/\phi_2^{(0)}\sim \sqrt{U}$ implies a nearly constant correction of the order of $\hbar$ for semiclassical states, where $U\approx \hbar^2/4$ remains asymptotically constant. The
first term in (\ref{phi2}) may also be relevant in intermediate regimes, where it would imply a $\delta\phi_2$ that behaves like $x^{1/4}$. The correction to
$\phi_2$ then increases asymptotically, unlike $\phi_3$, but less slowly than
$\phi_2^{(0)}$: we have $(\delta\phi_2)/\phi_2^{(0)}\sim x^{-3/4}$ from the first term in (\ref{phi2}).

\subsubsection{Lapse correction}

Given this solution for $\delta\phi_2$, we can go back to (\ref{p2linear}) as
a differential equation for $\delta N$. So far, we have not fully solved this equation and only used it to eliminate $\delta N$ from (\ref{Hsecond2}). Our solution for $\delta\phi_2$ obtained in this way now makes it possible to solve (\ref{p2linear}) for $\delta N$, although the lengthy form of (\ref{phi2}) makes it hard to find a complete analytical solution. Nevertheless, the form of the solution in certain limits will turn out to be instructive.

We first rewrite
equation (\ref{p2linear}) as
\begin{equation}
 0= \left(1-\frac{\mu}{x}\right)^{3/2} \left(-\frac{1}{2x^2\sqrt{1-\mu/x}}
   \:\delta\phi_2 + \left(\frac{\delta N}{\sqrt{1-\mu/x}}\right)'\right)
\end{equation}
such that
\begin{equation} \label{deltaNint}
 \delta N = -\frac{1}{2} \sqrt{1-\frac{\mu}{x}} \int \frac{\delta\phi_2}{
   x^2\sqrt{1-\mu/x}} {\rm d}x
\end{equation}
where (\ref{phi2}) should be inserted in the integral.
A simple integration is obtained in regimes in which both $C^2$-terms in
(\ref{phi2}) can be ignored, in which case
\begin{equation}
 \delta N \sim -\frac{2}{3} \frac{\sqrt{E}}{x^{3/4} (1-\mu/x)^{1/4}}+ F
 \sqrt{1-\frac{\mu}{x}}
\end{equation}
with a new integration constant $F$. The $F$-term just changes the background
lapse function by a constant factor $1+F$, which can be absorbed in the time
coordinate. The remaining correction to the lapse function,
\begin{equation} \label{deltaN}
 \delta N \sim -\frac{2}{3} \frac{\sqrt{E}}{x^{3/4} (1-\mu/x)^{1/4}}\,,
\end{equation}
shows an interesting asymptotic behavior of the correction which falls off
more slowly than the classical curvature correction $-\mu/x$ of the lapse
function. Using this term as a correction of Newton's potential in a
weak-field line element shows that non-local effects could imply larger
corrections than effective field theory in a derivative expansion, where the
leading correction would be of the order $1/x^3$
\cite{EffectiveNewton}. However, our simplified solution (\ref{deltaN}), based
on the $E$-term in (\ref{phi2}), does not apply in the completely asymptotic
regime where the $U$-term in (\ref{phi2}) would be dominant. Since this term, for asymptotically constant $U(x)$,
implies an asymptotic behavior of $\delta\phi_2\sim \sqrt{U}x$, the
corresponding $\delta N$ according to (\ref{deltaNint}) is $\delta N\sim
\sqrt{U} \log(\mu/x)$. Interpreted as a correction to Newton's potential, this
term suggests a relationship with infrared contributions, consistent with the
interpretation of fluctuation terms in quantum cosmological models that have the same origin as $U$ here \cite{Infrared,MiniSup}.

We are left with the equation $\dot{p}_3=0$, a differential equation for $L$.
It is straightforward to solve
\begin{eqnarray}
 \frac{\dot{p}_3}{\phi_3} &=& -\frac{\partial^2H}{\partial\phi_2^2} N^{(0)}+
 \left(\frac{\partial^2H}{\partial\phi_2\partial\phi_2'}N^{(0)}\right)'+
 \frac{U(x)\phi_2^{(0)}N^{(0)}}{\sqrt{\phi_1}\phi_3^4} -
 2\frac{\partial H}{\partial\phi_2} L+ \left(\frac{\partial
     H}{\partial\phi_2'}L\right)'\nonumber\\
&=& -\frac{1-\mu/x}{2x^2} +\frac{2U(x)}{C^4}
 \left(1-\frac{\mu}{x}\right)^6 - \frac{2\mu}{x^2}L+ \left(1-\frac{\mu}{x}\right)L'
\\ 
&=& \left(1-\frac{\mu}{x}\right)^3
\left(-\frac{1}{2(x-\mu)^2}+\frac{2U(x)}{C^4}\left(1-\frac{\mu}{x}\right)^3 +
  \left(\frac{L}{(1-\mu/x)^2}\right)'\right)=0\nonumber
\end{eqnarray}
for $L$, where we have used background solutions in all coefficients. The result is
\begin{eqnarray}
 &&L=-\frac{1-\mu/x}{2x}+ G\left(1-\frac{\mu}{x}\right)^2\\
&&- \frac{2U(x)x}{C^4}\left(1-\frac{\mu}{x}\right)^2
   \left(\left(1-\frac{\mu}{x}\right)^3+ \frac{3\mu}{2x}
   \left(1-\frac{\mu}{x}\right)^2+ \frac{3\mu^2}{x^2}\left(1-\frac{\mu}{x}\right)
 +\frac{3\mu}{x} \log\left(\frac{\mu}{x}\right)\right)\nonumber
\end{eqnarray}
with a new integration constant $G$. Since $L$ vanishes at $x=\mu$, the
evolution of fluctuations freezes at the horizon, just as the evolution of the
classical metric.

\subsubsection{Quantum effects}

We have obtained complete solutions, up to two remaining integrations. These are not only lengthy in analytical form but also require additional information about the function $U(x)$, which quantifies the strength of quantum effects. So far, we have mainly discussed $U$-dependent modifications in asymptotic low-curvature regimes, in which we assumed that $U(x)$ is nearly constant. The results were encouraging, in that they showed that a nearly constant $U$ also implies a nearly constant relative metric fluctuation, given by $\delta\phi_2/\phi_2^{(0)}$. Nevertheless, it is of interest to obtain independent information about the possible form of $U(x)$. 

Since the field $U(x)$ does not have a momentum, in the truncation to
second-order moments used here, it is not subject directly to an evolution
equation. (At higher moment orders, the uncertainty product
$\Delta(\phi_2^2)\Delta(p_2^2)-\Delta(\phi_2p_2)$, which equals $U$ to second
order, is not conserved. The momentum of $U$ can therefore be thought of as a
combination of higher-order moments that are eliminated in our truncation.)
However, it turns out that we can use another equation of motion in order to
derive a consistency condition for $U(x)$: We have implemented the leading
non-zero terms in the equation $\dot{p}_2=0$, which were of linear order in
$\delta\phi_2$ and $\delta N$. Since we used second-order constraints, there
is also a second-order contribution to $\dot{p}_2$. Setting this contribution
equal to zero for static solutions allows us to test the self-consistency of
the formalism. A long calculation (performed using Mathematica) implies an
equation for $U(x)$ of the form
\begin{eqnarray} \label{Ueq}
  0 &=& f_1(x)+f_2(x)U(x)+ f_3(x)U(x)^2+f_4(x)U'(x)\nonumber\\
    &&+ f_5(x) I[U]+ f_6(x)U(x)I[U]+ f_7(x)U'(x)I[U]\nonumber\\
  &&+ f_8(x)I[U]^2
\end{eqnarray}
where
\begin{equation}
    I[U]=\int \sqrt{x}(1-\mu/x)^{7/2} U(x) \, {\rm d}x
\end{equation}
and the $U$-independent coefficient functions are
\begin{eqnarray}
    f_1(x)&=&36C^8\sqrt{1-\frac{\mu}{x}}   \left(\sqrt{1-\frac{\mu}{x}} \left(1+\frac{3\mu}{2x}-\frac{3\mu^2}{2x^2}\right)+1\right)\\
    &&
-3 C^6E x^{1/2}\left(1-\frac{\mu}{x}\right) \left(\sqrt{1-\frac{\mu}{x}}\left(5+\frac{7\mu}{x}-\frac{6\mu^2}{x^2}\right)+9 \right)
+5 C^4 E^2  x \left(1-\frac{\mu}{x}\right)^{3/2}\nonumber\\
f_2(x)&=&
12 C^4x^2 \left(1-\frac{\mu}{x}\right)^{11/2} \left(\sqrt{1-\frac{\mu}{x}} \left(3+\frac{13\mu}{x}\right) - 1\right)\nonumber\\
&&-4C^2E x^{5/2} \left(1-\frac{\mu}{x}\right)^6 \left(\sqrt{1-\frac{\mu}{x}}\left(3+\frac{14\mu}{x}\right)  -1 \right)\\
f_3(x)&=& 16 x^{4} \left(1-\frac{\mu}{x}\right)^{11} \\
f_4(x)&=&16C^2x^3 \left(1-\frac{\mu}{x}\right)^7 \left(3 C^2-E \sqrt{x} \sqrt{1-\frac{\mu}{x}}\right)\\
f_5(x)&=& -6 C^4 x^{1/2} \left(1-\frac{\mu}{x}\right) \left(\sqrt{1-\frac{\mu}{x}}\left(5+\frac{7\mu}{x}- \frac{6\mu^2}{x^2}\right)+9\right)
+20 C^2 E x \left(1-\frac{\mu}{x}\right)^{3/2}\\
f_6(x)&=& -8 x^{5/2} \left(\sqrt{1-\frac{\mu}{x}}\left(3+\frac{14\mu}{x}\right)-1\right)  \left(1-\frac{\mu}{x}\right)^6\\
f_7(x)&=& -32 x^{7/2} \left(1-\frac{\mu}{x}\right)^{15/2} \\
f_8(x)&=& 20 x \left(1-\frac{\mu}{x}\right)^{3/2} \,.
\end{eqnarray}

\begin{figure}
    \centering
    \includegraphics[width=13cm]{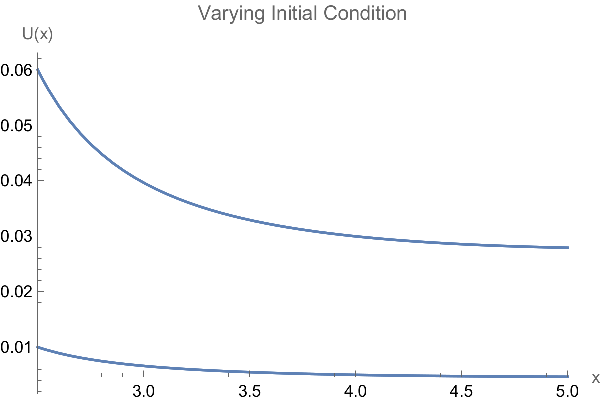}
    \caption{An example of two numerical results for $U(x)$ following from equation (\ref{Ueq}). Two choices of initial values were set for $U(x)$ at $x=2.5$, given by $U_{\rm in}=0.01$ in the lower curve and $U_{\rm in}=0.06$ for the upper curve, respectively. Constants are set as follows: $C=0.01, E=0.0001, \mu=1$. 
    \label{fig:Ux}}
\end{figure}

This long equation can be analyzed in the asymptotic regime if we assume that
$U(x)$ is of power-law form there. In the derivative terms, $xU'(x)$ is then
of the same order as $U(x)$, and asymptotically for $x\gg\mu$ with nearly
constant $U$ the integral behaves like $x^{3/2}$. In (\ref{Ueq}), the
contributions with coefficient functions $f_3(x)$, $f_6(x)$, $f_7(x)$, and
f$_8(x)$ are then dominant, such that the equation simplifies to
\begin{eqnarray}
&&  a U(x)^2+ b  U(x) I[U]/x^{3/2}\\
&&  +c xU'(x)I[U]/x^{3/2}+ d I[U]^2/x^3=0\nonumber 
\end{eqnarray}
with $x$-independent coefficients $a$, $b$, $c$ and $d$. For nearly constant $U$ at $x\gg\mu$,  we have $I[U]\sim \frac{2}{3}x^{3/2}U$, and therefore our equation takes the form 
\begin{equation}
    \tilde{a}U(x)^2+ c xU(x)U'(x)=0
\end{equation}
with a new constant $\tilde{a}$. The simplified equation therefore has solutions $U(x)=0$ or a power law for $U(x)$. Asymptotically, these solutions are consistent with our condition that $U(x)$ not be negative. Numerical solutions at smaller $x$, shown in Fig.~\ref{fig:Ux} confirm this behavior.

This result is encouraging because the non-negativity condition is motivated by the quantum-mechanics origin of our modifications, which is independent of the consistency conditions we checked for the constraint brackets. The observation that solutions respect the quantum condition indicates that the equations are self-consistent, not only as a model of modified gravity, but also from the perspective of quantum physics.

\section{Conclusions}

Any quantum theory, and in particular quantum gravity, is expected to imply
non-local behavior. Non-local action principles and their equations of motion are usually hard to solve, but if one assumes
a specific non-local action, it can
often be analyzed by mapping the theory to a local one in which classical degrees of freedom are coupled to auxiliary fields.  We have introduced here a new,
systematic quasiclassical formulation of spherically symmetric models in
quantum gravity with non-local corrections derived in a canonical
quantization. By implementing quantum fluctuations and correlations as
physical versions of what would usually be called auxiliary fields in a
non-local theory, a multi-field local theory is obtained in which coupling
terms are completely determined by the rules of canonical quantization.

The presence of new degrees of freedom implies that such quantum extended
theories are more complex than the classical model.  Working with vacuum
spherically symmetric models, we constructed a tractable constrained system in
which one of the metric components, $\phi_1$, is fixed by using the area radius
(a partial gauge fixing of the theory). Doubling the classical field content
by introducing second-order quantum moments, we therefore obtained a theory
for two independent fields that represent a single classical metric component
(the radial distance measure $\phi_2$) and its quantum fluctuation ($\phi_3$). While the
reduced system ignores cross-correlations between the radial distance
$\phi_2/\sqrt{\phi_1}$ and the area radius $\sqrt{\phi_1}$, it is formally consistent
for static solutions and allows explicit solutions in almost complete closed
form.

The fluctuation field $\phi_3$ couples dynamically to expectation value $\phi_2$, representing one of the metric components. The former field cannot vanish owing to uncertainty relations, and through the coupling terms it implies changes $\delta\phi_2$ of the metric field compared with its classical behavior. Through canonical equations of motion, the staticity condition determines the lapse function $N$ for a given $\phi_2$, such that $\delta N$ inherits certain changes from $\delta\phi_2$. Using the appearance of these fields in a classical-type line element, we obtain a quantum-corrected space-time geometry from
\begin{eqnarray} \label{line}
    {\rm d}s^2&=& -(N+\delta N)^2{\rm d}t^2+ \frac{(\phi_2+\delta\phi_2)^2}{x^2}{\rm d}x^2+ x^2 {\rm d}\Omega^2\\
    &\sim& -(N^2+2N\delta N){\rm d}t^2+ \frac{\phi_2^2+2\phi_2\delta\phi_2}{x^2}{\rm d}x^2+ x^2{\rm d}\Omega^2\nonumber
\end{eqnarray}
to first order in $\delta N$ and $\delta\phi_2$. The latter values are given by the rather lengthy expressions  (\ref{deltaNint}) and  (\ref{phi2}), respectively. However, a word of caution is in order when we organize our solutions in this form: So far, we have checked the consistency of our quasiclassical constraints only for static configurations, and therefore we can use a line element of the form (\ref{line}) only for static slicings. It might be tempting to apply a more general coordinate transformation once solutions have been put into the form of a line element, but by doing so we would leave the range of validity of our derivations here. The cosmological analysis \cite{Inhom} extended our static constraints to non-static ones, observing that consistency then requires an inclusion also of fluctuations of $\phi_1$. An application to black-hole models remains to be completed.

We have observed several interesting features of our solutions. In particular,
the quasiclassical approximation breaks down before the horizon is reached,
which suggests that non-local effects may be crucial for horizon dynamics of
quantum black holes. A confirmation of this expectation would, however, have
to await a solution of higher-order quasiclassical approximations, as well as an extension to non-static configurations that would allow us to use different space-time slicings.  

The asymptotic behavior is more reliable within the restrictions of our model. We analyzed it by studying solutions for one of the new quantum fields that corresponds to the uncertainty of a state in quantum mechanics. For this field, we found an asymptotic fall-off behavior consistent with a positivity condition. Our quasiclassical solutions are therefore consistent with the existence of an underlying quantum state of static, spherically symmetric space-times. In a full quantum field theory, important properties such as positivity would be implied by unitary evolution. The fact that we observed a positivity property without explicitly deriving unitary evolution from the quasiclassical constraints indicates that our treatment is self-consistent and does reveal features of an underlying quantum theory of gravity. Our analysis therefore shows that quasiclassical methods are promising in applications
to inhomogeneous models of quantum gravity. They allow explicit derivations of quantum corrections without requiring additional assumptions beyond what is provided by canonical quantization.

\section*{Acknowledgements}

We thank Unnati Akhouri, Joseph Balsells, Aurora Colter, Jackson Henry, and
Jakub Mielczarek for discussions and comments on this paper.  This work was
supported in part by NSF grants PHY-1912168 and PHY-2206591, a Chateaubriand
Fellowship, and an RI NASA Space Grant.

%\bibliographystyle{../preprint}
%\bibliography{../Bib/QuantGra,../Bib/Tunneling}

\end{document}